%% file: main.tex
\documentclass[10pt]{article}
\usepackage{fullpage}
\usepackage{graphicx}
\usepackage{subcaption}
\usepackage{xcolor} 

\usepackage{hyperref}

\hypersetup{
    colorlinks=true,
    linkcolor=blue,
    citecolor= blue,
    filecolor=magenta,      
    urlcolor=magenta,
    allbordercolors={1 1 1}
    }
\usepackage{natbib}
\setcitestyle{authoryear,round}

\input{math_commands}

\usepackage{amsmath}
\usepackage{amssymb}
\usepackage{mathtools}
\usepackage{amsthm}
\usepackage{mathrsfs}

\usepackage[nameinlink, capitalize, noabbrev]{cleveref}

\theoremstyle{plain}
\newtheorem{theorem}{Theorem}[section]
\newtheorem{proposition}[theorem]{Proposition}

\theoremstyle{definition}

\theoremstyle{remark}

\usepackage[font=small,labelfont=bf]{caption}

\title{\textbf{ReLUs Are Sufficient for Learning Implicit Neural Representations}}

\author{
Joseph Shenouda\\
University of Wisconsin-Madison\\
\texttt{jshenouda@wisc.edu} \\
\and
Yamin Zhou \\
University of Wisconsin-Madison\\
\texttt{zhou463@wisc.edu}\\
\and
Robert D. Nowak\\
University of Wisconsin-Madison\\
\texttt{rdnowak@wisc.edu}\\
}
\date{}

\begin{document}

\maketitle

\begin{abstract}
Motivated by the growing theoretical understanding of neural networks that employ the Rectified Linear Unit (ReLU) as their activation function, we revisit the use of ReLU activation functions for learning implicit neural representations (INRs). Inspired by second order B-spline wavelets, we incorporate a set of simple constraints to the ReLU neurons in each layer of a deep neural network (DNN) to remedy the spectral bias. This in turn enables its use for various INR tasks. Empirically, we demonstrate that, contrary to popular belief, one \emph{can learn} state-of-the-art INRs based on a DNN composed of only ReLU neurons. 
Next, by leveraging recent theoretical works which characterize the kinds of functions ReLU neural networks learn, we provide a way to quantify the regularity of the learned function. 
This offers a principled approach to selecting the hyperparameters in INR architectures.
We substantiate our claims through experiments in signal representation, super resolution, and computed tomography, demonstrating the versatility and effectiveness of our method. The code for all experiments can be found at \url{https://github.com/joeshenouda/relu-inrs}.
\end{abstract}

\section{Introduction}

\begin{figure}[ht]
    \centering
    \begin{subfigure}[t]{0.45\columnwidth}
        \centering
         \includegraphics[width=\linewidth]{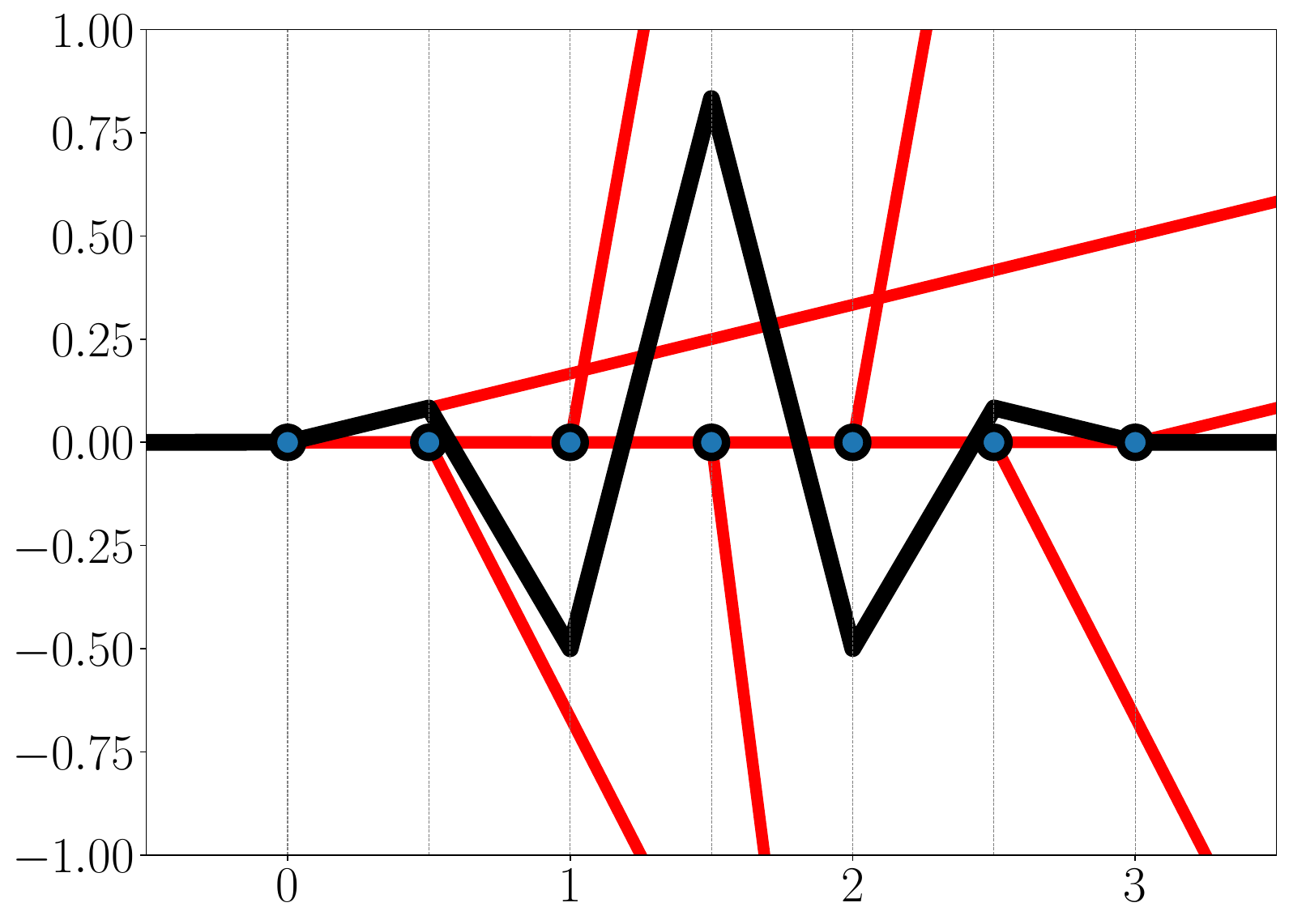}
         \caption{}
         \label{fig:bspline-w-act}
     \end{subfigure}
     \hfill
     \begin{subfigure}[t]{0.45\columnwidth}
        \centering
         \includegraphics[width=\linewidth]{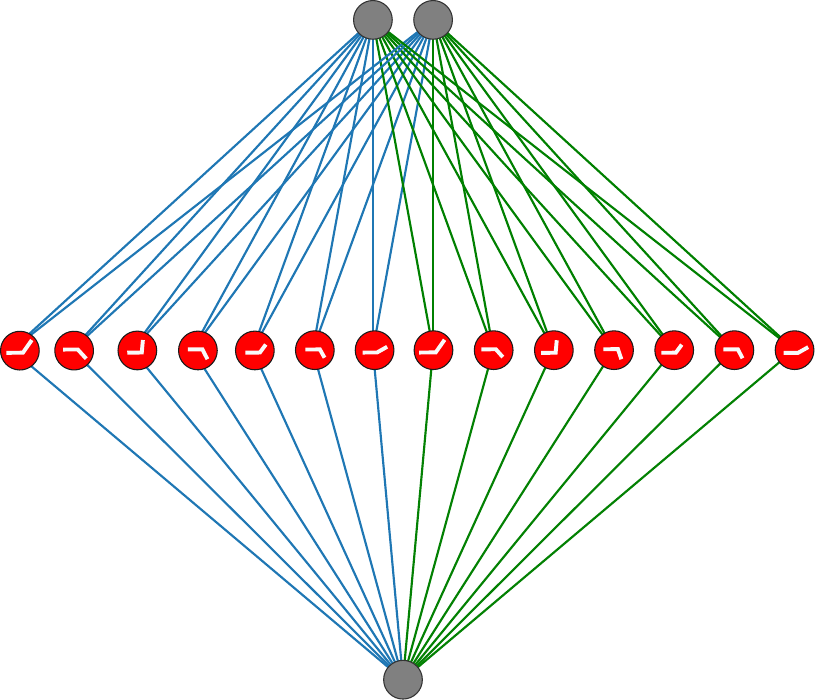}
        \caption{}
        \label{fig:bspline-w-net-as-relu}
     \end{subfigure}
    \caption{(Left) A plot of the second order B-spline wavelet activation function. The red lines indicate the seven non-local ReLU functions that make up a single second order B-spline wavelet, shown in black. 
    (Right) A BW-ReLU neural network with two neurons represented as a constrained ReLU network with 14 neurons. Within each group the orientation of each ReLU relative to the others is fixed. The input and output weights are learned and shared across groups of neurons. Shared input/output weights are denoted by the same color.}
    \label{fig:bspline-w-as-relus}
\end{figure}
Recently, training deep neural networks (DNNs) to learn implicit neural representations (INRs) has led to advancements in various vision-related tasks. These include, but are not limited to, computer graphics \citep{mildenhall2021nerf}, image processing \citep{chen2021learning}, and signal representation \citep{sitzmann2020implicit}. They have also shown great promise in biomedical imaging, where they can be used for sparse-view computed tomography (CT) \citep{sun2021coil, wu2023self}. Many INR tasks involve learning continuous representations of images, unlike image classification tasks which train DNNs on high-dimensional data. INRs learn a continuous representation of an image by training a DNN on the low-dimensional coordinates of the image.
For such imaging tasks, the success of the INR hinges on the ability of the DNN to efficiently approximate and learn high-frequency components of the image. This has, thus far, prohibited the use of DNNs with ReLU activations as they have been shown to exhibit a \emph{spectral bias} \citep{rahaman2019spectral}—an inherent bias of ReLU DNNs which causes them to struggle in approximating high-frequency functions when trained via gradient descent. 

Therefore, in order to circumvent this problem while still utilizing the power of neural networks, practitioners in the INR community have adopted preprocessing techniques \citep{mildenhall2021nerf, tancik2020fourier} and many non-standard activation functions. These include, but are not limited to, the sine function \citep{sitzmann2020implicit}, the Gaussian function \citep{ramasinghe2022beyond}, or the complex Gabor wavelet \citep{saragadam2023wire}. However, using these highly unorthodox activation functions has raised many questions about the theoretical properties of INRs \citep{yuce2022structured}.
Therefore, motivated by the fact that a majority of our theoretical understanding of neural networks is based on standard ReLU DNNs, we revisit the use of ReLU activations for learning INRs.

Consider a shallow ReLU neural network of the form
\begin{align}\label{eq:shallow-relu-net}
    f_{\bm{\theta}}(\vx) = \sum^{K}_{k=1} v_k \sigma(\vw^{T}_k \vx - b_k)
\end{align}
where $\sigma(z) = \text{max}\{0,z\}$.
The neural network function is a linear combination of a set of ``atomic functions"-in this case the ReLU neurons $\{ \sigma(\vw^{T}_k \vx - b_k) \}^{K}_{k=1}$. The monotonically increasing nature of the ReLU activation function implies that each ReLU neuron contributes globally to the function and is highly coherent with the others.  
When attempting to fit this network to data, if the weights of one neuron changes, all the other neurons in the layer must also adjust to correct for the influence of that one neuron. This makes the function highly sensitive to even small changes in the parameters, and results in a severely ill-conditioned optimization problem. This ill-conditioning mandates an inordinate number of iterations when training the network using gradient-type optimization methods, even when using modern optimizers such as Adam. This problem is exacerbated in INR tasks where the data is densely sampled, low dimensional and exhibits high frequencies.

Guided by this simple insight, our investigation takes a different approach than previous works and focuses solely on remedying the optimization problem. This is in contrast to wholly replacing the activation functions of the neurons which generate our features.
Specifically, we examine whether a localized ReLU-based activation function can effectively address and overcome the ill-conditioning. To do this, we propose incorporating constraints to groups of ReLU neurons within each of the hidden layers in a ReLU DNN. For each group, all of the neurons share the same weights. 
Our constraints ensure that every seven ReLU neurons in the hidden layer is effectively applying the activation function $\psi(x): \R \to \R$ where,
\begin{align}\label{eq:bspline-as-relu}
\begin{split}
\psi(x) &= \frac{1}{6}(\sigma(x) + \sigma(x-3)) -\frac{16}{3} \sigma\left(x- 1.5\right)- \frac{8}{6}\left(\sigma\left(x-0.5\right)+\sigma\left(x- 2.5\right)\right) +\frac{23}{6} \left(\sigma(x-1) +  \sigma(x-2)\right).
\end{split}
\end{align}
This activation function corresponds to a second order B-spline wavelet introduced in \citet{chui1992compactly, unser1993family}. A plot of the second order B-spline wavelet in the univariate case is shown in \Cref{fig:bspline-w-act}. For ease of exposition we will refer to neural networks that use this activation function simply as BW-ReLU neural networks.

Thus, when training a BW-ReLU neural network of the form
\begin{align}\label{eq:bspline-wavelet-nn}
    g_{\bm{\theta}}(\vx) = \sum_{k=1}^{K} v_k \psi( \vw^{T}_k \vx - b_k),
\end{align}
we are still learning a function which can be exactly represented by a ReLU neural network with $7K$ neurons as demonstrated in \Cref{fig:bspline-w-net-as-relu}. At an intuitive level, the compact nature of these neurons allows for each neuron to fit different parts of the function without affecting the contributions from other neurons. In \Cref{sec:ill-conditioning-relus} we provide a more thorough discussion on the ill-conditioning and how the unique properties of the BW-ReLU can remedy it.
Our experiments in \Cref{sec:experiments} demonstrate the effectiveness of this approach on various INR tasks. 

Next, having developed a method for learning ReLU-based INRs, we leverage the recently developed theory characterizing the kinds of functions ReLU neural networks learn when fit to data \citep{savarese2019infinite, ongie2020function, parhi2021banach, shenouda2023vector}. We show how our BW-ReLU neural network functions fit into this mathematical framework and how we can measure the \emph{variation norm} of these functions. This provides a measure of the regularity of the learned function. We also give new insights into how this regularity is effected when one reparameterizes the INR with a scaling parameter $c > 0$ such that the neurons are defined as $v\cdot\psi(c\cdot(\vw^{T}\vx - b))$. This heuristic is employed in all of the activation functions introduced for INRs despite being poorly understood~\citep{sitzmann2020implicit, ramasinghe2022beyond, saragadam2023wire}. Moreover, we show how the variation norm of the BW-ReLU neural network provides a good indication for how well the network can generalize to unseen data. This suggests a principled way to tune INRs without the need for an additional validation dataset.
In summary, our contributions are:
\begin{enumerate}
    \item \textbf{A method for learning ReLU-based INRs}: By incorporating a simple set of constraints on the neurons of a ReLU neural network, we can overcome the ill-conditioning inherent to ReLU networks. We demonstrate that this approach can be used in multiple INR tasks and performs comparably to other INR architectures that use unconventional activation functions.
    \item \textbf{Insights on INR generalization}: We present a way to measure the regularity of BW-ReLU neural networks by leveraging recent theoretical results on the kinds of functions ReLU neural networks learn. This regularity is measured in terms of the variation norm. We discuss how this perspective provides insights into some of the heuristics employed in learning INRs and how BW-ReLU neural networks with a lower variation norm tend to generalize better.
\end{enumerate}

\section{Related Works}
The ReLU is perhaps the simplest activation function utilized in DNNs. Thus much of the recent DNN theory has focused on this setting \citep{bach2017breaking,savarese2019infinite, arora2018understanding,ongie2020function,parhi2021banach}. They are also useful in practice as they induce sparse activation which can be exploited for compression or speeding up inference \citep{li2023lazy,kurtz2020inducing, mirzadeh2024relu}.

However, neural networks with ReLU activations are typically not utilized for INR tasks due to their \emph{spectral bias} \citep{rahaman2019spectral}. To remedy this, pre-processing techniques \citep{tancik2020fourier, mildenhall2021nerf}
and unconventional activation functions \citep{sitzmann2020implicit, ramasinghe2022beyond, saragadam2023wire} have been used alongside or instead of traditional ReLU DNNs. Our results brings into question the necessity of these unorthodox approaches by showing that a ReLU-based DNN can be trained for various INR tasks.

Our approach is inspired by B-spline wavelets which were first developed and studied in \cite{chui1992compactly, unser1993family, unser1992asymptotic}.
Moreover, our BW-ReLU neural networks are very related to the concept of ridgelets which were originally developed and studied in \cite{candes1998ridgelets, candes1999ridgelets}.

\section{Remedying the Spectral Bias of ReLU Neural Networks}\label{sec:ill-conditioning-relus}
\begin{figure}
  \centering
  \includegraphics[width=\linewidth]{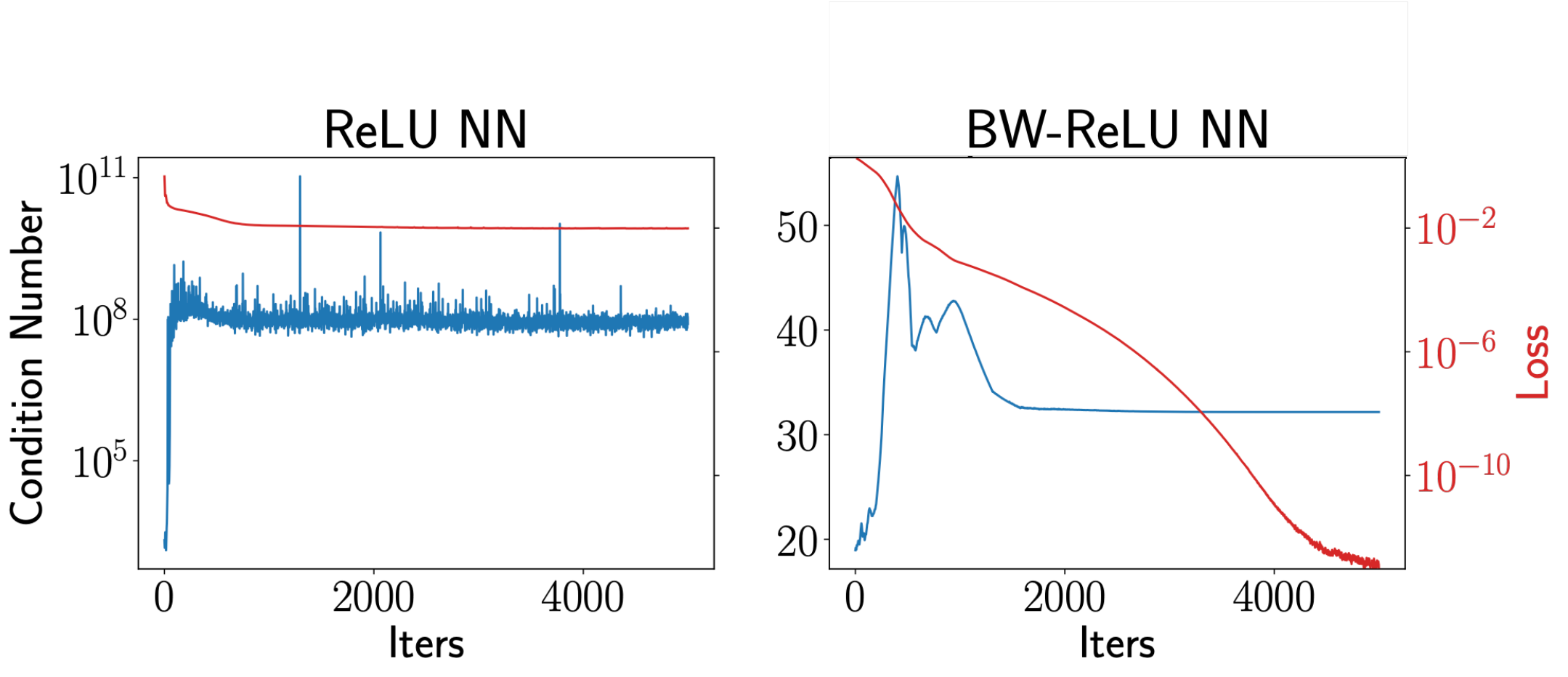}
  \caption{Condition number of feature embedding matrix generated by ReLU vs. BW-ReLU neural networks during training. We see that the ReLU produces a severely ill-conditioned feature matrix at initialization and throughout training. In contrast, the BW-ReLU neural network enjoys a very well conditioned feature matrix all throughout training. The rate of convergence is also correlated with how well conditioned the feature matrix in both cases.
  }
  \label{fig:cond-num-univariate}
\end{figure}

In this section, we provide a more thorough discussion on why ReLU neural networks tend to exhibit a spectral bias when trained to fit low-dimensional datasets. We formalize the intuitions presented in the introduction and explain how the strong coherence between ReLU neurons results in an ill-conditioned optimization problem, which significantly slows down convergence. We also discuss key properties of the second-order B-spline wavelet, which remedy this ill-conditioning (and consequently the spectral bias) making them suitable for INR tasks.

Consider approximating a univariate function $u:D \to \R$ by a univariate ReLU neural network $f: D \to \R$ of the form
\begin{align}
    f(x) &= \sum_{k=1}^{K} v_k \sigma(w_kx - b_k). 
\end{align}
Where $x \in D,\quad w_k \in \{-1,1\}$ and $ v_k,b_k,c \in \R$. Here $D=[r_1, r_2]$ is a bounded domain, $K$ denotes the width of the network and $\sigma:\R \to \R$ denotes the ReLU activation function defined as $\sigma(\cdot) = \max\{0, \cdot\}$. The restriction of the input weights $w_k$ to $\{-1,1\}$ follows from the homogeneity of the ReLU (i.e., for any $\alpha > 0$ we have that $\sigma(\alpha z) = \alpha\sigma(z)$). Now due to the fact that $\sigma(z) = z - \sigma(-z)$ we can further restrict the input weights to be $+1$ by introducing a skip connection

\begin{align}
    f(x) = c + a x + \sum_{k=1}^{K} v_k \sigma(x - b_k) 
\end{align}
where $ \quad x \in D$ and  $v_k,b_k,c,a \in \R $. By only considering the approximation on the domain $D$, setting $c = u(r_1)$ and $b_k \in [-1,1]$ the network can be further reduced to 
\begin{align}
    f_{\bm{\theta}}(x) = c  + \sum_{k=1}^{K} v_k \sigma(x - b_k)
\end{align}
where $\bm{\theta} = (v_k, b_k)^{K}_{k=1}$ denotes the parameters of the network.

Now suppose we train the neural network to approximate $u$ by sampling the function and training the network on a univariate dataset $(x_i, y_i)^{N}_{i=1}$ to minimizing the $\ell_2$ loss. This corresponds to solving the following optimization problem
\begin{align}\label{opt:univariate-shallow-opt}
    \min_{\bm{\theta}} \| \mathbf{\Phi}^{T}\vv - \vy \|^2_2 =     \min_{\bm{\theta}} \vv^{T} \mathbf{\Phi} \mathbf{\Phi}^{T} \vv - 2 (\mathbf{\Phi}\vy)^{T}\vv,
\end{align}
where $\vy \in \R^{N}$ is a vector containing the labels for all $N$ samples and $\vv \in \R^{K}$ is a vector containing the output weight of each neuron. Moreover, $\mathbf{\Phi} \in \R^{K \times N}$ is the feature embedding matrix, which depends on the input biases and is learned throughout training of the network,
\begin{align}
    \mathbf{\Phi}_i = \begin{bmatrix}
        \sigma(x_1 - b_i), \cdots, \sigma(x_N - b_i) \\
    \end{bmatrix}.
\end{align}
Neural networks typically solve \eqref{opt:univariate-shallow-opt} via gradient descent. If we consider the gradient step update on just the output weights $\vv$ then it is clear from \eqref{opt:univariate-shallow-opt} that this is equivalent to taking a gradient descent step on the least squares problem over a fixed set of features $\mathbf{\Phi}$.
The effectiveness of each gradient step to minimize the objective is dependent on the condition number of the Hessian, $\mathbf{\Phi}\mathbf{\Phi}^{T}$\citep{boyd2004convex}. For a real symmetric matrix $A \in \R^{m \times m}$ the condition number is defined as $\kappa(A) = \frac{\lambda_1}{\lambda_m}$ where $\lambda_1$ and $\lambda_m$ are the largest and smallest eigenvalues respectively.

In the context of least squares, if the matrix $\mathbf{\Phi}\mathbf{\Phi}^{T}$ has a very high condition number then the curvature of the loss landscape (with respect to the output weights $\vv$) will vary dramatically in different directions. This hinders the effectiveness of each gradient step and results in requiring many more gradient steps to converge.

When using ReLU activations, the feature embedding matrix $\mathbf{\Phi}\mathbf{\Phi}^{T}$ is typically severely ill-conditioned which results in a slow converge. This ill-conditioning of the features is present at initialization and persists throughout training, see for example \Cref{fig:cond-num-univariate}.  To understand why, consider two ReLUs with biases that are close to each other. In this case they are nearly colinear (causing ill-conditioning).  On the other hand, if all the neurons are orthogonal to each other (impossible with ReLUs), then $\mathbf{\Phi}\mathbf{\Phi}^{T}$ would be perfectly well conditioned.

To understand this quantitatively, consider approximating the univariate function $u: [-1,1] \to \R$ with a ReLU neural network using a fixed set of neurons. Instead of minimizing the $\ell_2$ loss over a set of finite samples we will instead consider minimizing the $L_2$ loss between the neural network $f_{\bm{\theta}}$ and the function $u$
\begin{align}
    \min_{\vv} \frac{1}{2} \int^{1}_{-1} \left(u(x) - u(-1) - \sum_{k=1}^{K} v_k \sigma(x - b_k) \right)^2 dx.
\end{align}
A simple expansion shows that solving this optimization problem is equivalent to solving 
\begin{align}
    \min_{\vv} \vv^{T} \mathbf{G}_{\sigma} \vv - \vr^{T}_{u,\sigma} \vv. 
\end{align}
Where $\vr_{u,\sigma} \in \R^{K}$ is defined as
\begin{align}
    (\vr_{u,\sigma})_i = \int^{1}_{-1} (u(x) - u(-1)) \sigma(x - b_i).
\end{align}
Moreover, $\mathbf{G}_{\sigma} \in \R^{K \times K}$ is the Gram matrix for the feature embedding matrix and is defined as,
\begin{align}
    \mathbf{G}_{i,j} := \int^{1}_{-1} \sigma(x - b_i) \sigma(x-b_j) dx.
\end{align}
Note that this is analogous to $\mathbf{\Phi} \mathbf{\Phi}^{T}$ discussed above. Therefore, the condition number of $\mathbf{G}_{\sigma}$ determines how ill-conditioned our problem is and indicates how effective each gradient step will be. Theorem 4 in
\citet{zhang2023shallow} quantifies the condition number of $\mathbf{G}_{\sigma}$.

\begin{theorem}[\cite{zhang2023shallow}]
Suppose $\{b_j\}^{K}_{j=1}$ are quasi-evenly spaced on $[-1,1]$, $b_j = -1 +\frac{2(j-1)}{K} + o(\frac{1}{K})$. Let $\lambda_1 \geq \lambda_2 \geq \cdots \geq \lambda_K \geq 0$ be the eigenvalues of the Gram matrix  $\mathbf{G}_{\sigma}$, then the condition number of $\mathbf{G}_{\sigma}$ satisfies 
\begin{align*}
    \kappa(\mathbf{G}_{\sigma}) = \lambda_1 /\lambda_K = \Omega(K^3).
\end{align*}
\end{theorem}
The theorem shows us that even when the ReLU neurons are maximally separated, solving \eqref{opt:univariate-shallow-opt} for the output weights is a severely ill-conditioned problem. Moreover, the condition number of the feature matrix grows at a cubic rate. Thus, as we increase the number of neurons in the network, which increases the approximation power of the model, we are simultaneously hindering the ability of gradient descent to optimize the model.

We can potentially remedy this ill-conditioning by instead considering second order B-spline wavelets as our activation function \eqref{eq:bspline-as-relu} and approximating $u$ by the following BW-ReLU neural network,
\begin{align}
    g_{\bm{\theta}}(x) = \sum_{k=1}^{K} v_k \psi(w_k x - b_k).
\end{align}
As discussed earlier this is equivalent to incorporating constraints on a regular ReLU neural network such that each group of ReLU neurons share weight and have a fixed orientation relative to the other neurons in the group. We first present a simple proposition establishing that any ReLU neural network can be represented by a BW-ReLU neural network over a bounded domain.
\begin{proposition}\label{lemma:bspline_w_reps_relu}
   Let $f: [-1,1] \to \R$ denote a ReLU neural network with $K$ neurons. The network is of the form
    \begin{align}
        f(x) = c + \sum_{k=1}^{K} v_k \sigma(w_k x-b_k)
    \end{align}
    where $v_k \in \R$, $b_k \in [-1,1]$, and $w_k \in \{-1,+1\}$ are the parameters of the model. Then there exists a BW-ReLU neural network $g: \R \to \R$ with the same number of neurons of the form 
    \begin{align}
        g(x) = c + \sum_{k=1}^{K} 24 v_k \psi\left(\frac{1}{4} (w_k x - b_k)\right)
    \end{align}
    such that $g$ represents $f$ on the bounded domain $[-1,1]$.
\end{proposition}
The proof is in \cref{sec:appendix_a}. The proposition establishes the fact that on a bounded domain (the setting relevant to INRs) any function which we can represent using a ReLU neural network can also be represented by a BW-ReLU neural network with the same number of neurons. Therefore, no representation power is lost by incorporating these constraints into the ReLU network.

Now consider approximating the function $u: [-1,1] \to \R$ by a univariate BW-ReLU neural network optimizing over the output weights with a fixed set of neurons. The $L_2$ loss in this case is,
\begin{align}
    \min_{\vv} \frac{1}{2} \int^{1}_{-1} \left(u(x) - \sum_{k=1}^{K} v_k \psi( w_k x - b_k)\right)^2 dx.
\end{align}
This is equivalent to solving
\begin{align}
    \min_{\vv} \vv^{T} \mathbf{G}_{\psi} \vv - \vr^{T}_{u,\psi} \vv
\end{align}
where
\begin{align}
    (\mathbf{G}_{\psi})_{i,j} = \int^{1}_{-1} &\psi(w_i x - b_i)\psi(w_j x - b_j) dx. 
\end{align}
Our next theorem shows that the Gram matrix in this setting is far better conditioned than in the case of ReLUs.
\begin{theorem}\label{thm:bspline_w_cond}
    Consider a BW-ReLU neural network with  $K=2^{J}-1$ neurons of the form,
    \begin{align}\nonumber
    \begin{split}
        \psi_{j,k}(x) = 2^{j/2} \psi\left(2^{j}\frac{3}{2}(x+1) - k\right) &\quad j = 0,\dots, J-1 \\
        &\quad k=0, \dots, 2^{j}-1,
    \end{split}
    \end{align}
    for any $J \in \mathbb{N}^+$. For each scale $j = 0,\cdots, J-1$ we have $k = 0,...,2^{j}-1$ shifted versions of the B-spline wavelets. 
    Let $\lambda_1 \geq \lambda_2 \cdots \geq \lambda_K \geq 0$ be the eigenvalues of $\mathbf{G}_{\psi}$.
    Using these neurons the condition number of $\mathbf{G}_{\psi}$ satisfies
    \begin{align}
        \kappa(\mathbf{G}_{\psi}) = \lambda_1 / \lambda_K = \mathcal{O}(1).
    \end{align}
\end{theorem}

The proof is in \Cref{sec:appendix_b}. We note that the normalization constant is not required and merely simplifies the analysis. The proof relies on key properties of the B-spline wavelets. In particular, the fact that they are \emph{semiorthogonal}. This ensures that wavelets of different scale are orthogonal to each other. For instance in the dyadic wavelet system developed in the theorem  $\langle \psi_{j,k}, \psi_{i,\ell} \rangle = 0$ for any $i \neq j$ and any $k,\ell \in \mathbb{Z}$.

In \Cref{fig:cond-num-univariate} we present a numerical example of how the condition number of the feature embedding matrix changes during training when using a ReLU or BW-ReLU neural network to fit a univariate function. This illustrates how the Gram matrix of the feature embeddings remain well conditioned both at initialization and throughout training for the BW-ReLU neural network while the ReLU neural network suffers from a poorly conditioned feature embedding matrix all throughout training. We also see that the well conditioned feature embedding matrix correlates with a faster rate of convergence.
Moreover, our experiments in \Cref{sec:experiments} also demonstrate that deep BW-ReLU neural networks are not susceptible to this ill-conditioning and can be utilized for real INR tasks. 

The Gaussian \cite{ramasinghe2022beyond} and Gabor wavelets \cite{saragadam2023wire} have also been utilized as activation functions for INR tasks due to their localized nature. However, there is little theory characterizing the kinds of functions such networks learn and the effects their hyperparameters have on the learned function. In the next section we leverage the fact that we are ultimately learning a ReLU neural network. This allows us to utilize much of the recent theory characterizing of the function space associated with ReLU neural networks giving insights into some of the heuristics used when training INRs. 

\section{Variation Norm and Scale}

\begin{figure}
    \centering
        \begin{subfigure}[b]{0.3\columnwidth}
         \centering
         \caption*{$\|g_{\theta}\|_{\mathcal{V}} = 131.34$}
         \includegraphics[width=\columnwidth]{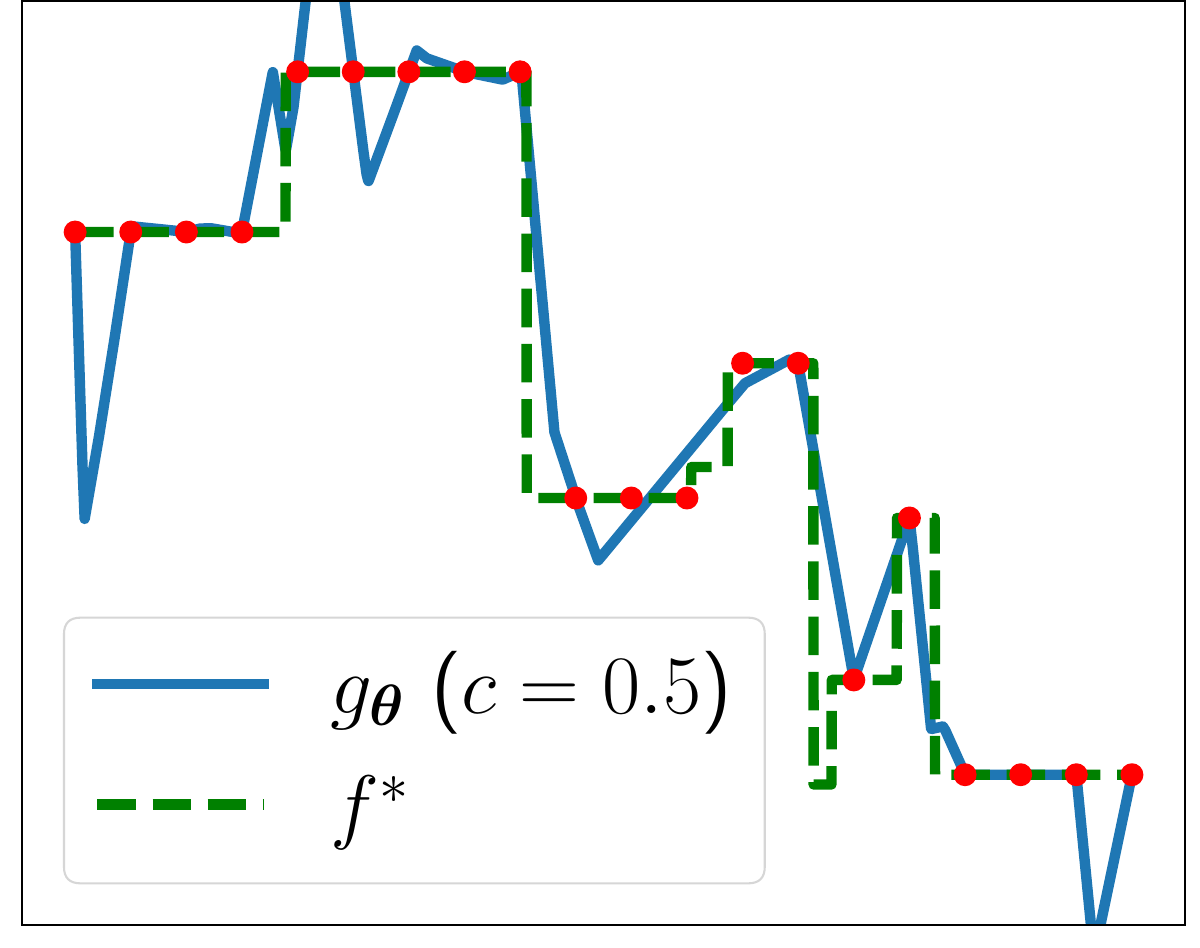}
         \caption{}
     \end{subfigure}
    \hfill
     \begin{subfigure}[b]{0.3\columnwidth}
         \centering
         \caption*{$\|g_{\theta}\|_{\mathcal{V}} = 17.76$}
         \includegraphics[width=\columnwidth]{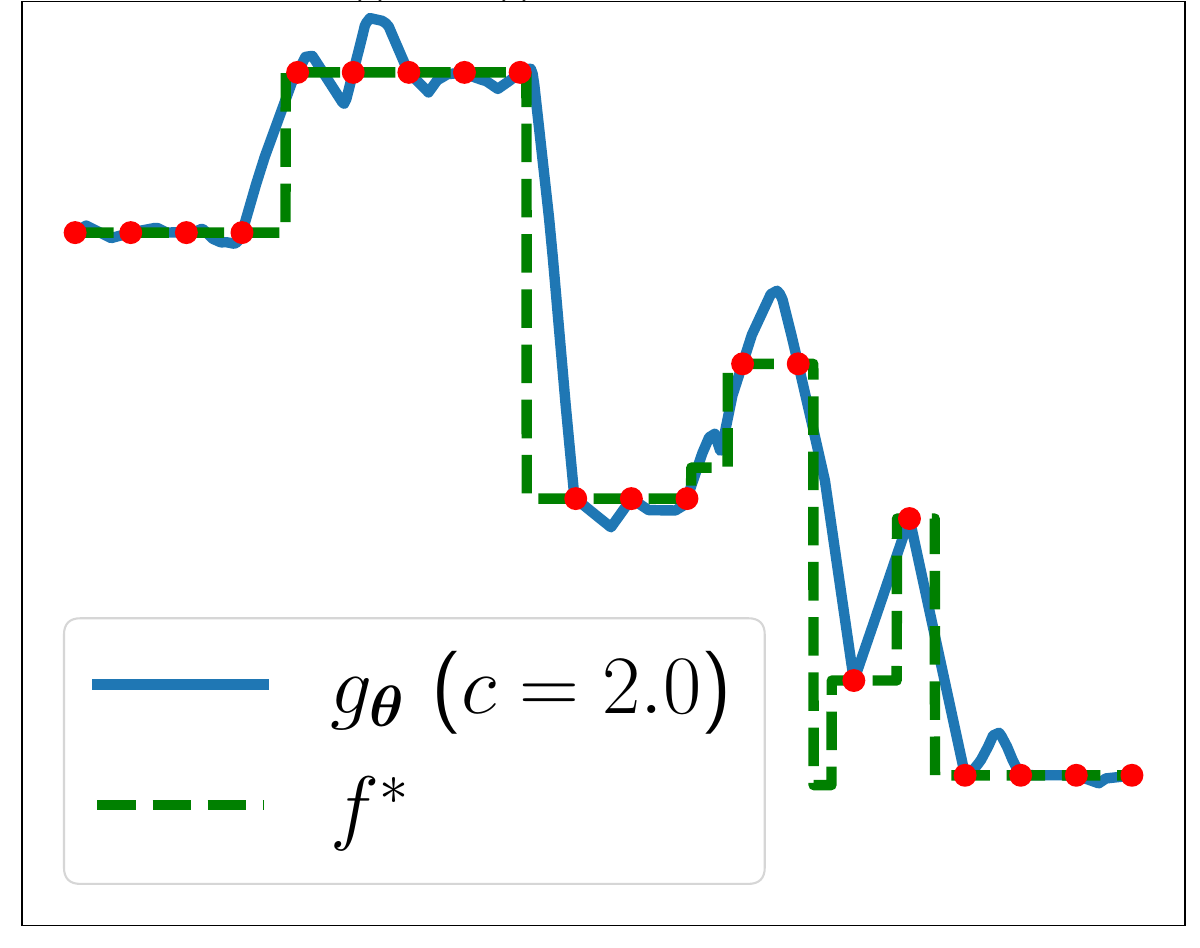}
         \caption{}
     \end{subfigure}
     \hfill
     \begin{subfigure}{0.3\columnwidth}
         \centering
         \caption*{$\|g_{\theta}\|_{\mathcal{V}} = 103.12$}
         \includegraphics[width=\columnwidth]{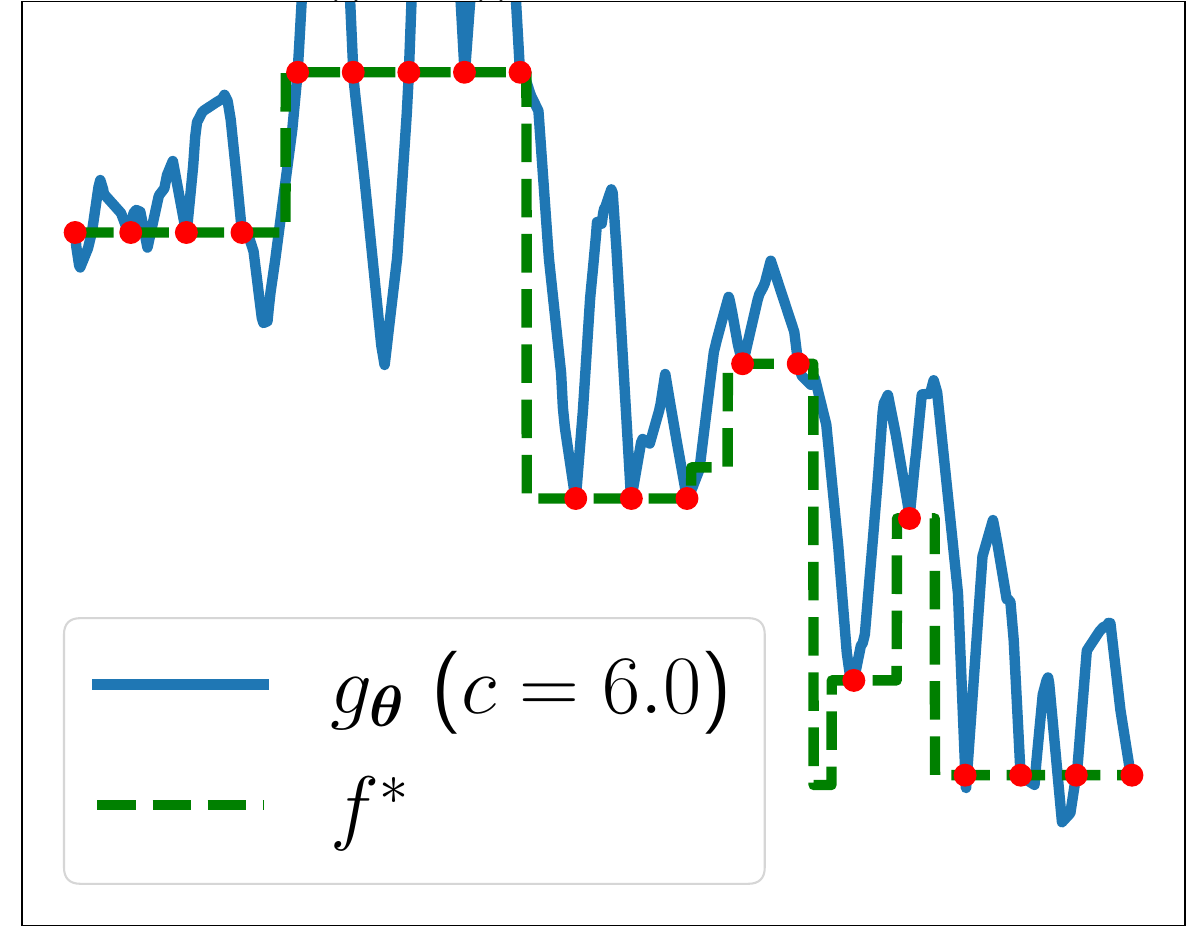}
         \caption{}
     \end{subfigure}
    \caption{The variation norm of BW-ReLU neural networks, $g_{\boldsymbol{\theta}}$ with various scales trained on univariate data. The red dots indicate our samples from the ground truth function $f^{*}$. We see that making $c$ too low leads to a poor fit to the data and a very high variation norm. On the other hand making $c$ to large results in a very oscillatory fit to the data. The interpolator which generalizes best corresponds to the one with the lowest variation norm.}
    \label{fig:tv_scale}
\end{figure}
Having demonstrated how we \emph{can} learn ReLU-based INRs by imposing constraints on groups of neurons, we now explain the benefits of obtaining such a representation. Due to the prevalence and simplicity of the ReLU activation function, much of our theoretical understanding of DNNs has been focused on those with ReLU activations. In particular, a line of work \cite{bach2017breaking, savarese2019infinite, ongie2020function, parhi2021banach, wojtowytsch2020banach, parhi2023deep, siegel2023characterization, bartolucci2023understanding, chen2023multi, shenouda2023vector, zeno2023how} has investigated the \emph{kinds} of functions which are learned when fitting ReLU neural networks to data. This mathematical framework has provided many insights into the inner workings and success of neural networks. 

For instance, it sheds light on why neural networks seemingly break the curse of dimensionality \citep{klusowski2018approximation} and how they differ from kernel methods \citep{parhi2023near}. Moreover, this perspective offers an explanation for the role of various heuristics that are commonly employed when training neural networks such as weight decay, bottleneck linear layers and skip connections \citep{parhi2022kinds, shenouda2023vector}. We will soon see how this perspective can also provide insights into one of the hyperparameters uniquely used when learning INRs.

These efforts have revealed that for neural networks with ReLU activations the common regularization technique of weight decay corresponds to regularizing a certain \emph{variation norm} \citep{kurkova2001bounds}. This norm is related to total variation and is the appropriate norm for functions  represented by ReLU neural networks. This norm provides a measure of smoothness for the learned function and, remarkably, can be easily computed in terms of the weights of the model \citep{neyshabur2015path, savarese2019infinite, parhi2021banach}. 
In the univariate case, the variation norm is the total variation of the derivative of the function $f_{\bm{\theta}}$ defined by the neural network.  If $f_{\bm{\theta}}$ has a continuous derivative, then this is equivalent to the $L_1$ norm of the second derivative. The variation norm is also well-defined for continuous functions having discontinuous derivatives (in which case the second derivative will be a generalized function having Dirac impulses). In the multivariate case,  the variation norm is essentially the $L_1$ norm of the Radon transform of the Laplacian (second derivative operator) of the function. This is equivalent to considering the $L_1$ norm of the second derivative of the function along each direction of the multidimensional domain. For a ReLU neuron, the directional second derivative is $0$ in all but one direction determined by the orientation of the neuron. And in that direction the second derivative is an impulse with magnitude equal to the slope of the ReLU.

Thus, the variation norm has a clear connection to the smoothness of the function, as measured by the size of the second derivatives. We refer the reader to the recent survey \citet{parhi2023deep} for more details.

We now present this variation norm and show how it can be computed for our BW-ReLU neural networks.

First, consider a function $f_{\bm{\theta}}$ represented by a ReLU neural network of the form,
\begin{align}\label{eq:shallow-net}
    f_{\bm{\theta}}(\vx) = \sum^{K}_{k=1} \vv_k \sigma(\vw^{T}_k \vx - b_k),
\end{align}
where $\bm{\theta} = (\vv_k, \vw_k, b_k)^{K}_{k=1}$ and $\sigma(\cdot): \R \to \R$ is the ReLU activation function applied elementwise. Each neuron, $\eta_{\vv, \vw} (\vx) = \vv \sigma(\vw^{T}\vx - b)$
contributes to the variation norm of the function. Explicitly, the variation norm for each ReLU neuron is,
\begin{align}
    \| \eta_{\vv, \vw} \|_{\mathcal{V}} = \| \vv\|_2 \|\vw\|_2,
\end{align}
where $\|\cdot\|_{\mathcal{V}}$ denotes the variation norm of a function. Note the biases are not regularized. The norm of the entire function is computed by summing up the contribution from each neuron such that,
\begin{align}
    \|f_{\bm{\theta}}\|_{\mathcal{V}} = \sum_{k=1}^{K} \| \vv_k \|_2 \| \vw_k \|_2.
\end{align}
Now consider a function represented by a BW-ReLU neural network of the form
\begin{align}
    g_{\bm{\theta}}(\vx) = \sum_{k=1}^{K} \vv_k \psi(\vw^{T}_k \vx - b_k).
\end{align}
This network can be exactly represented by a ReLU neural network with $7K$ neurons by the definition of the B-spline wavelet activation function \eqref{eq:bspline-as-relu}. 

For each BW-ReLU neuron $\gamma_{\vw,\vv}(\vx) = \vv \psi(\vw^{T}\vx - b)$ its variation norm is found by summing up the variation norm of each of the scaled ReLUs in the definition of the B-spline wavelet \eqref{eq:bspline-as-relu}, from this we get that
\begin{align}
    \| \gamma_{\vw, \vv} \|_{\mathcal{V}} = 16 \|\vv\|_2 \|\vw\|_2.
\end{align}
We can see that it is simply a multiple of the total variation of each ReLU neuron. Again, since the variation norm of a neural network function consists of summing up the variation norm of each neuron we have that for the BW-ReLU neural network its variation norm can be explicitly computed as,
\begin{align}
    \|g_{\bm{\theta}}\|_{\mathcal{V}} = 16 \sum_{k=1}^{K} \|\vv\|_2 \|\vw\|_2.
\end{align}

\subsection{Understanding Scaling via Regularization}
It is common to introduce an additional hyperparameter to the neurons used in INR applications, which scales each activation function as follows.  Consider an activation function $\zeta$.  Each neuron applies this activation function to $\vw^T\vx-b$, producing the output $\zeta(\vw^T\vx-b)$.  Most INR networks employ an additional activation \emph{scaling} parameter $c>0$, so that each neuron output is instead computed by $\zeta(c(\vw^T\vx-b))$.  The scaling parameter can have a significant impact for certain activation functions, while being ineffectual others.  

To illustrate, first consider the ReLU activation.  In this case, $\sigma(c(\vw^T\vx-b)) = c\sigma(\vw^T\vx-b)$, so the the scaling affects the magnitude of output but has no other effect.  This is due to the fact that the ReLU activation is homogeneous (linear activations and leaky ReLUs are also homogeneous in this way).  However, the activations commonly used in INR applications are inhomogeneous.  Take for example, the commonly used sine activation function $\sin(\vw^{T}\vx-b)$ \citep{sitzmann2020implicit}.  In this case if we scale by $c$, then $\sin(c(\vw^T\vx-b))$ has the same output range/magnitude but will have faster or slower oscillations relative to  $\sin(\vw^T\vx-b)$, if $c>1$ or $c<1$.  

Including the scaling parameter with inhomogeneous activations functions, like the sine function, can change the shape and variations of the activation function.  Activation functions like the Gaussian \citep{ramasinghe2022beyond} the complex Gabor wavelets \citep{saragadam2023wire} and the B-spline wavelet used here, are compressed or dilated depending on whether $c>1$ or $c<1$, which also makes the support of the activation function smaller or larger, accordingly.  These observations illustrate why scaling the activation function may have significant effects on INR performance. Generally speaking, large values of $c$ tend to increase oscillations or variations in the activation function (and possibly decrease the support of the activation function).  This fact has been used to explain why large values of $c$ may help capture more detailed structure in INR applications \citep{yuce2022structured}.

The story, however, is a bit subtle.  Consider a standard training problem of the form
\begin{align}
\min_{\bm{\theta}} \sum_{i=1}^N \ell\Big(\vy_i,\sum^{K}_{k=1} \vv_k \zeta\big(c(\vw_k^T\vx_i-b_k)\big)\Big), \
\end{align}
where the minimization is with respect to all the weights and biases. If we place no restrictions on the allowable ranges of the input weights and biases, then the scaling factor $c$ can simply be absorbed into the weights and biases.  So why does $c$ play a crucial role?  One answer is that the neural networks are trained via gradient descent methods, typically initialized with small random weights.  This, coupled with the fact that the training objective is nonconvex, tends to favor solutions that fit the data with weights of small magnitudes (even though the data might also be fit using much larger weights) \citep{vardi2021implicit}.  This is sometimes referred to as the implicit regularization of gradient descent. 

To understand why the scaling factor $c$ can play a significant role, it is enlightening to consider explicit regularization in the form of the commonly used weight decay regularization term, which is proportional to the sum of squared weights in the network.  This is supported by established theoretical connections between weight decay and implicit regularization \citep{chizat2020implicit}. The overall training objective is to minimize the sum of losses plus the weight decay regularization term
\begin{align}
\sum_{i=1}^N \ell\Big(\vy_i,\sum^{K}_{k=1} \vv_k \zeta\big(c(\vw_k^T\vx_i-b_k)\big)\Big) + \lambda \sum^{K} 
 _{k=1} \|\vv_k\|_2^2 + \|\vw_k\|_2^2,
 \end{align}
where $\ell$ is a loss function and $\lambda>0$ is the weight decay parameter.
We can simply reparameterize the problem by absorbing $c$ into the weights to obtain an equivalent optimization
\begin{align}
\sum_{i=1}^N \ell\Big(\vy_i,\sum^{K}_{k=1} \vv_k \zeta(\vw_k^T\vx_i-b_k)\Big) + \lambda \sum^{K}_{k=1} \|\vv_k\|_2^2 + \frac{\|\vw_k\|_2^2}{c^2}.
\end{align}
Both objectives have the same global minima.  The second objective clearly reveals the effect of the scaling factor.  If $c>1$, then there is less regularization applied to the input weights compared to the output weights, and vice-versa if $c<1$.  So if $c>1$, then the regularizer encourages solutions with \emph{larger} input weights and hence increased oscillations or variations in inhomogeneous activation functions, like the sine, Gabor, or B-spline wavelet activations.  In the case of localized activations like the wavelets, larger values of $c$ also reduce the support (spatial scale) of the activation functions. However, unlike other activation functions, for B-spline wavelets the scaling parameter's effect on the regularity of the function can be readily understood in terms of the variation norm. Let us reparameterize our BW-ReLU neural network with a fixed scale $c > 0$ such that,
\begin{align}
    g_{\bm{\theta}}(\vx) = \sum_{k=1}^{K} \vv_k \psi(c \cdot (\vw^{T}_k \vx - b_k)).
\end{align}
 It follows from the previous discussion that the variation norm of the function represented by this neural network can be computed as,
\begin{align}\label{eq:variation_norm_bspline_c}
    \|g_{\bm{\theta}} \|_{\mathcal{V}} = 16c \sum_{k=1}^{K} \|\vw_k\|_2 \|\vv_k\|_2.
\end{align}
Thus a very high $c$ result in functions that is more irregular while using lower values of $c$ can lead to smoother functions.

In \Cref{fig:tv_scale} we present a simple univariate data fitting problem to illustrate the role of the scaling parameter $c$ and how it  can effect regularity of the learned function. We see that the interpolator which generalizes best is the one that minimizes the variation norm. Using a $c$ value which is too large results in a very oscillatory function with a high variation norm. However, if we instead make $c$ too small then the output weight must increase considerably to compensate for the wider B-spline wavelets. Moreover, in \Cref{exp:ct-path-norm} we illustrate how the variation norm can be a good indicator for how well our learned function will perform when using INRs to solve inverse problems.

\section{Experiments}\label{sec:experiments}
Here we demonstrate how our BW-ReLU neural networks can be as effective as other INR architectures for three INR tasks. We compared our method against SIREN \citep{sitzmann2020implicit}, WIRE \citep{saragadam2023wire} and ReLUs + Positional Encoding (P.E.) introduced in \cite{mildenhall2021nerf}. The positional encoding is a pre-processing technique which maps the low dimensional coordinates to higher dimensional Fourier features. The hyperparameters for each of the INR architectures were tuned to to give the best results. For all of our experiments we utilized a three hidden layer DNN and trained with the Adam optimizer. The full training details for all experiments can be found in \Cref{sec:appendix-c}. Moreover, all the code for reproducing the experiments can be found at \url{https://github.com/joeshenouda/relu-inrs}.

\begin{figure*}
     \centering
      \begin{subfigure}[b]{0.3\columnwidth}
         \centering
         \includegraphics[width=\columnwidth]{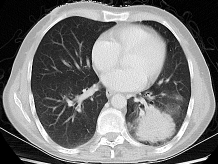}
         \caption{Original}
     \end{subfigure}
     \hfill
     \begin{subfigure}[b]{0.3\columnwidth}
         \centering
         \caption*{\textbf{31.5 dB \tiny($\pm 0.311$) }}
         \includegraphics[width=\columnwidth]{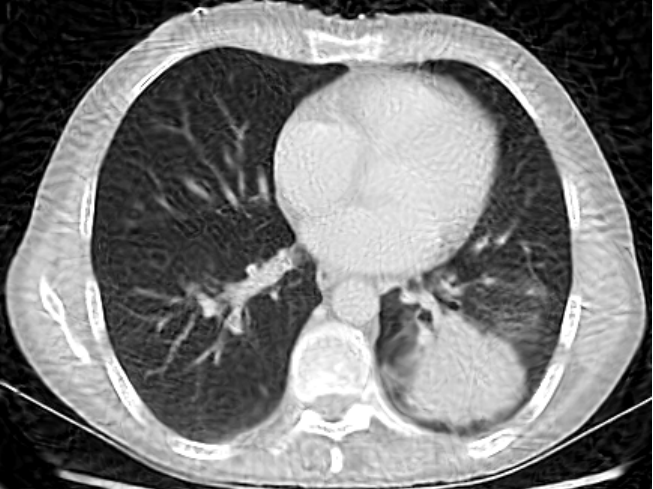}
         \caption{BW-ReLU}
     \end{subfigure}
     \hfill
     \begin{subfigure}[b]{0.3\columnwidth}
         \centering
         \caption*{31.16 dB \tiny($\pm 0.196$)}
         \includegraphics[width=\columnwidth]{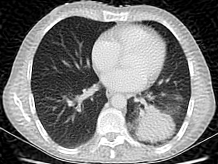}
         \caption{SIREN}
     \end{subfigure}
     \hfill
      \begin{subfigure}[b]{0.3\columnwidth}
         \centering
         \caption*{30.47 dB \tiny{($\pm 0.153$)}}
         \includegraphics[width=\columnwidth]{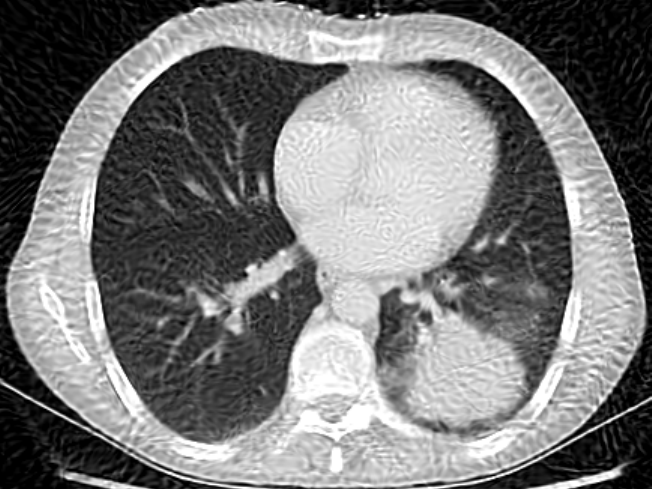}
         \caption{WIRE}
     \end{subfigure}
     \hfill
     \begin{subfigure}[b]{0.3\columnwidth}
         \centering
         \caption*{27.5 dB \tiny{($\pm 0.059$)}}
         \includegraphics[width=\columnwidth]{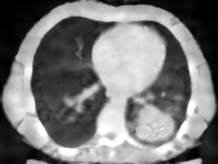}
         \caption{ReLU+P.E.}
     \end{subfigure}
     \hfill
     \begin{subfigure}[b]{0.3\columnwidth}
         \centering
         \includegraphics[width=\columnwidth]{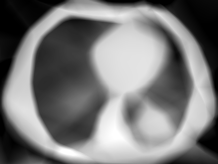}
         \caption{ReLU}
     \end{subfigure}
    \caption{Experiments on computed tomography reconstruction with various INR architectures. We report average PSNR and standard error across five random trials.}
    \label{fig:cond_act_func_synth}
\end{figure*}
\subsection{Computed Tomography(CT) Reconstruction}
In this experiment we simulated CT reconstruction by taking $100$ equally spaced CT measurements of a $326 \times 435$ chest X-ray image \citep{clark2013cancer}. \Cref{fig:cond_act_func_synth} shows the results compared to other INR architectures showing that our BW-ReLU neural networks perform just as well and perhaps slightly better than conventional INR architectures. Full experiment details can be found in \Cref{app:ct_dets}.

\subsection{Signal Representation}
\begin{figure*}
     \centering
     \begin{subfigure}[b]{0.3\columnwidth}
         \centering
         \includegraphics[width=\columnwidth]{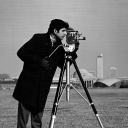}
         \caption{Original}
     \end{subfigure}
     \hfill
     \begin{subfigure}[b]{0.3\columnwidth}
         \centering
         \caption*{107.9 dB \tiny{($\pm 1.077$)}}
         \includegraphics[width=\columnwidth]{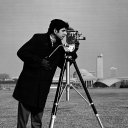}
         \caption{BW-ReLU}
         \label{fig:superres-bspline-w}
     \end{subfigure}
     \hfill
    \begin{subfigure}[b]{0.3\columnwidth}
         \centering
         \caption*{105.9 dB \tiny{($\pm 0.836$)}}
         \includegraphics[width=\columnwidth]{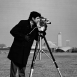}
         \caption{SIREN}
     \end{subfigure}
     \hfill
     \begin{subfigure}[b]{0.3\columnwidth}
         \centering
         \caption*{87.2 dB \tiny{($\pm 6.102$)}}
         \includegraphics[width=\columnwidth]{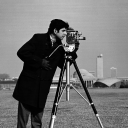}
         \caption{WIRE}
         \label{fig:superres-wire}
    \end{subfigure}
     \hfill
    \begin{subfigure}[b]{0.3\columnwidth}
         \centering
         \caption*{23.82 dB \tiny{($\pm 0.774$)}}
         \includegraphics[width=\columnwidth]{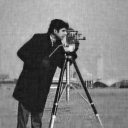}
         \caption{ReLU + P.E.}
     \end{subfigure}
    \hfill
    \begin{subfigure}[b]{0.3\columnwidth}
         \centering
         \includegraphics[width=\columnwidth]{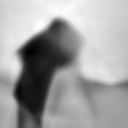}
         \caption{ReLU}
         \label{fig:superres-relu}
     \end{subfigure}
    \caption{Experiments on signal representation for various INR architectures. We report average PSNR and standard error across five random trials.}
    \label{fig:signal_rep}
\end{figure*}

In this experiment we fit the four INR architectures to the standard $256 \times 256$ cameraman image. The results are shown in \Cref{fig:signal_rep} where we have also included the case of using a traditional ReLU DNN for comparison. The results show that our method can perform comparably to the newly introduced INR architectures achieving a high PSNR as fast as the other methods. Full experiment details can be found in \Cref{app:sigrep_dets}.

\subsection{Super resolution}
\begin{figure*}[!th]
     \centering
     \begin{subfigure}[b]{0.3\columnwidth}
         \centering
         \includegraphics[width=\columnwidth]{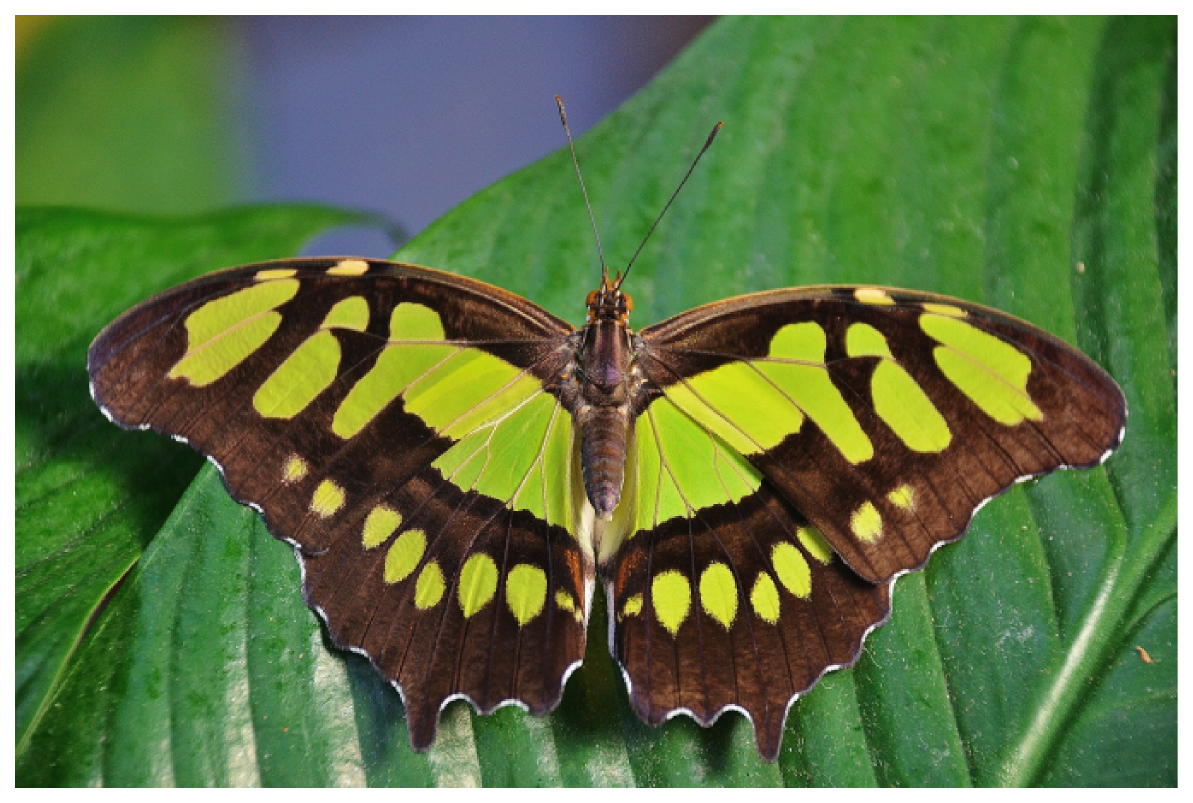}
         \caption{Original}
     \end{subfigure}
     \hfill
     \begin{subfigure}[b]{0.3\columnwidth}
         \centering
         \caption*{\textbf{27.1 dB \tiny($\pm 0.013$)}}
         \includegraphics[width=\columnwidth]{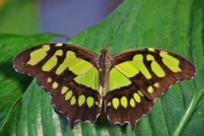}
         \caption{BW-ReLU}
         \label{fig:superres-bspline-w}
     \end{subfigure}
     \hfill
    \begin{subfigure}[b]{0.3\columnwidth}
         \centering
         \caption*{26.3 dB \tiny($\pm 0.212$)}
         \includegraphics[width=\columnwidth]{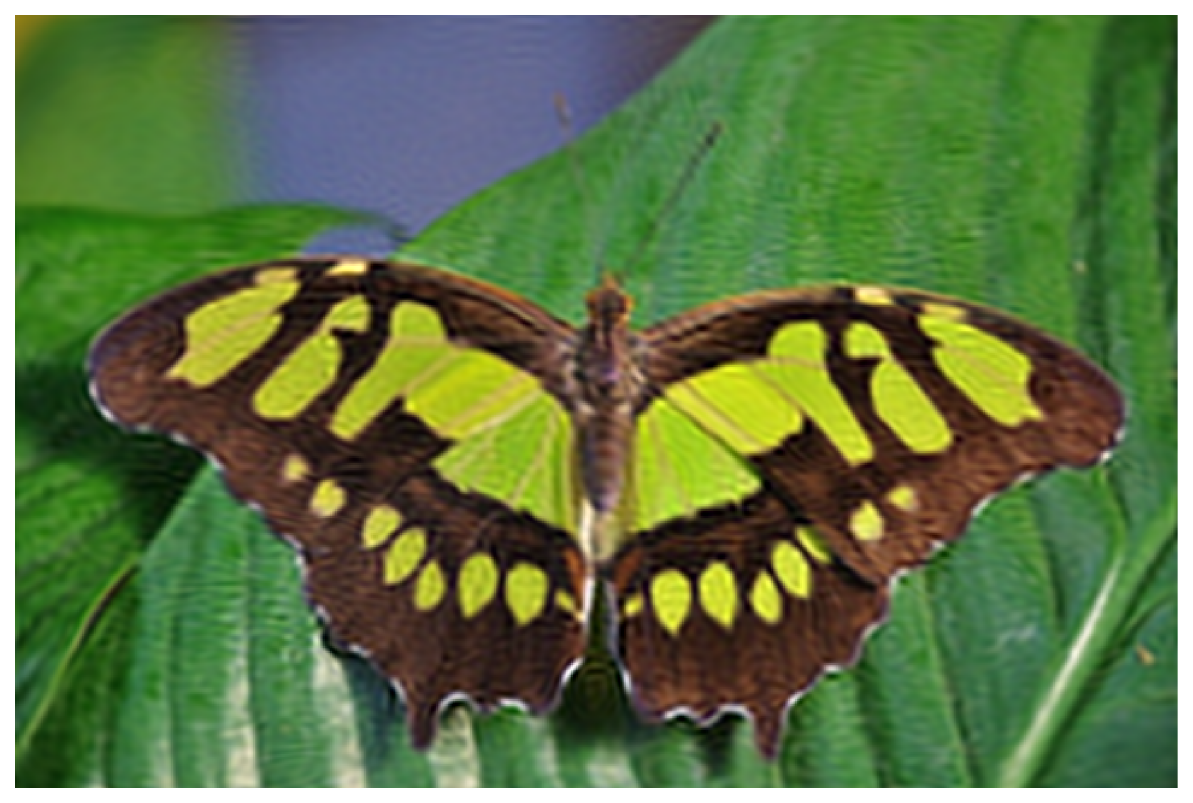}
         \caption{SIREN}
     \end{subfigure}
     \hfill
     \begin{subfigure}[b]{0.3\columnwidth}
         \centering
         \caption*{27.05 dB \tiny($\pm 0.048$)}
         \includegraphics[width=\columnwidth]{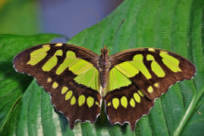}
         \caption{WIRE}
         \label{fig:superres-wire}
    \end{subfigure}
     \hfill
    \begin{subfigure}[b]{0.3\columnwidth}
         \centering
         \caption*{26.2 dB \tiny($\pm 0.083$)}
         \includegraphics[width=\columnwidth]{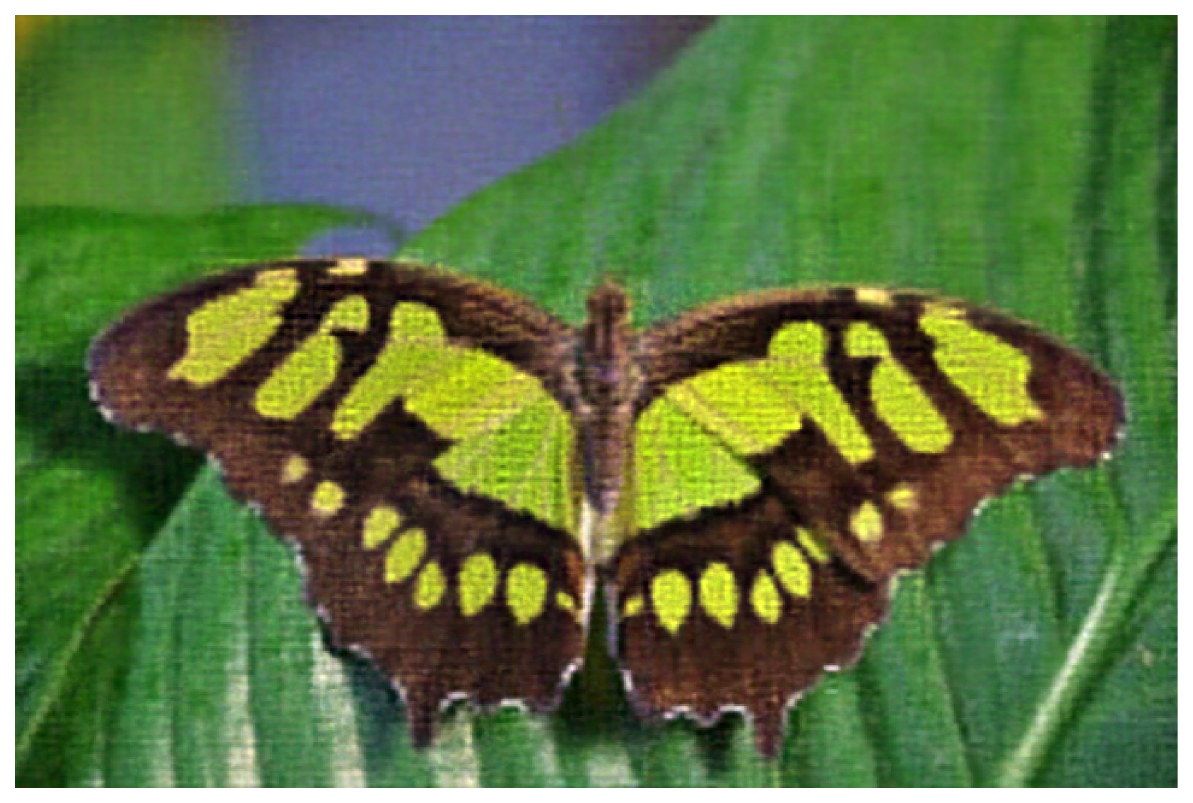}
         \caption{ReLU+P.E.}
     \end{subfigure}
    \hfill
    \begin{subfigure}[b]{0.3\columnwidth}
         \centering
         \includegraphics[width=\columnwidth]{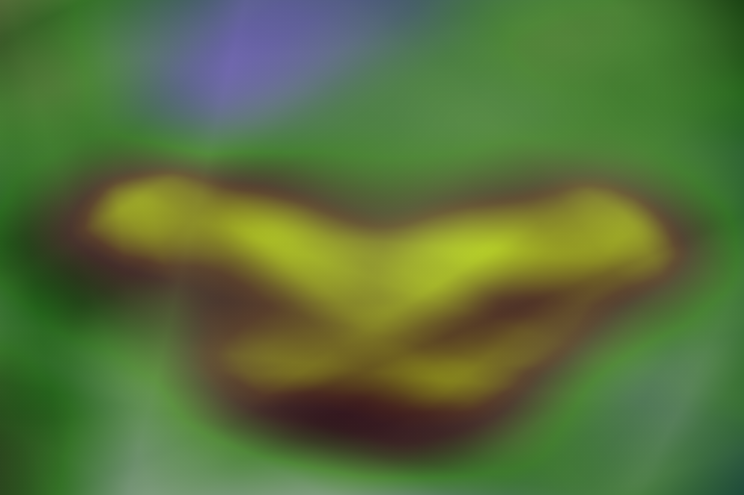}
         \caption{ReLU}
         \label{fig:superres-relu}
     \end{subfigure}
    \caption{Experiments on the super resolution task for various INR architectures. We report average PSNR and standard error across five random trials.}
    \label{fig:superres-butterfly}
\end{figure*}

We implemented a $4\times$ super resolution on an image from the DIV2K image dataset \citep{agustsson2017ntire}. Again we see that in this case our BW-ReLU neural networks performs on par and slightly better than the rest.
 The results on all four INR architectures are shown in \Cref{fig:superres-butterfly}. Full experiment details can be found in \Cref{app:supres_dets}.

\subsection{Low Norm Solutions and Inverse Problems}\label{exp:ct-path-norm}
\begin{figure*}[!h]
     \centering
         \begin{subfigure}[b]{0.4\columnwidth}
         \centering
         \caption*{$\sum_{\ell} \|g^{\ell}_{\bm{\theta}}\|_{\mathcal{V}} = 8650$}
         \includegraphics[width=\columnwidth]{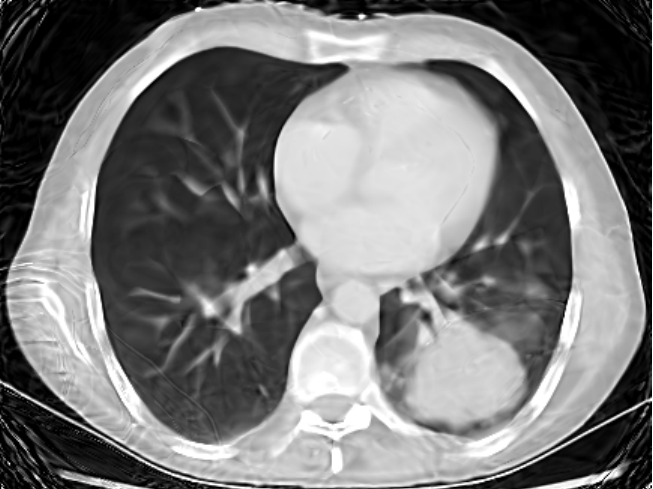}
         \caption{PSNR: 29.9 dB\\
         $c=1$}
     \end{subfigure}
     \hfill
     \begin{subfigure}[b]{0.4\columnwidth}
         \centering
         \caption*{$\sum_{\ell} \|g^{\ell}_{\bm{\theta}}\|_{\mathcal{V}} = 1978$}
         \includegraphics[width=\columnwidth]{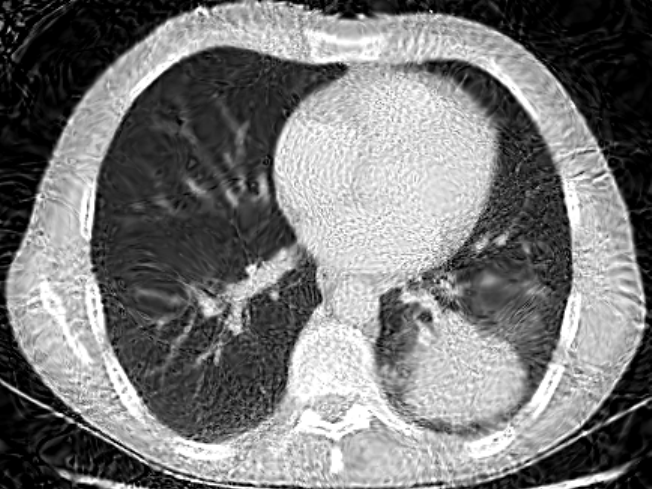}
         \caption{PSNR: 28.7 dB\\
         $c=2$}
     \end{subfigure}
     \hfill
     \begin{subfigure}[b]{0.4\columnwidth}
         \centering
         \caption*{$\sum_{\ell}\|g^{\ell}_{\bm{\theta}}\|_{\mathcal{V}} = 1292$}
         \includegraphics[width=\columnwidth]{figures_ct/brighter_bw_relu_best_ct32_dB.pdf}
         \caption{PSNR: \textbf{32.1 dB}\\
         $c=3$}
     \end{subfigure}
     \hfill
      \begin{subfigure}[b]{0.4\columnwidth}
         \centering
         \caption*{$\sum_{\ell} \|g^{\ell}_{\bm{\theta}}\|_{\mathcal{V}} = 1571$}
         \includegraphics[width=\columnwidth]{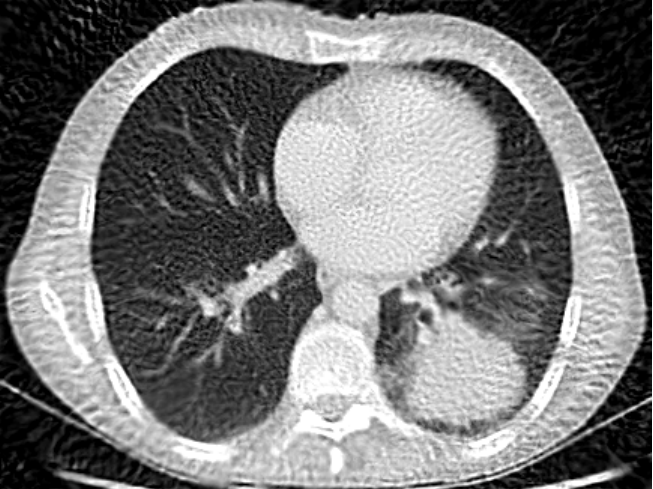}
         \caption{PSNR: 29.1 dB\\
         $c=5$
         }
     \end{subfigure}
    \caption{Four BW-ReLU DNNs trained on the CT reconstruction task with different values of $c$. All networks are trained to the same training loss. We see that the $c$ which produces the highest PSNR corresponds to the one with the lowest variation norm across all layers.}
    \label{fig:ct_path_norm}
\end{figure*}

For our last experiment we trained the BW-ReLU neural network for the CT reconstruction task with three different scaling parameters. In \Cref{fig:ct_path_norm} all three BW-ReLU neural networks that were used to reconstruct the image achieved the same training loss on the CT measurement data. However we see that the model with the smallest variation norm across all 3 layers corresponds to the best reconstruction. This suggests a principled methodology for choosing the scaling parameter $c$ in the case of inverse problems.

\section{Conclusion}
In this work we presented a simple way to utilize ReLU DNNs for INR tasks. Unlike previous works we focusing solely on remedying the ill-conditioning of the optimization problem without sacrificing the ReLU by drawing a connection to B-spline wavelets. We then related our methodology to the function space associated with ReLU neural networks and showed how this framework can be useful in understanding and quantifying the regularity of our INRs. This connection  suggests a more principled approach to tuning INRs.
For future work it would interesting to apply this technique to more INR tasks. In particular, neural radiance fields and physics informed neural networks.

\section*{Acknowledgement}
The authors would like to thank Liu Yang and Rahul Parhi for helpful conversations. The authors would also like to thank Ahmet Alacaoglu for his insights in the later stages of this project. Joseph Shenouda was supported by the ONR MURI grant N00014-20-1-2787 and NSF grant DMS-2023239. Robert Nowak was supported in part by the NSF
grants DMS-2134140 and DMS-2023239, the ONR MURI grant N00014-20-1-2787, and the AFOSR/AFRL grant FA9550-18-1-0166.

\section*{Broader Impact}
This paper presents work whose goal is to advance the field of Machine Learning. There are many potential societal consequences of our work, none which we feel must be specifically highlighted here.

\bibliography{refs}
\bibliographystyle{abbrvnat}

\newpage
\appendix
\onecolumn
\section{Proof of Proposition  \ref{lemma:bspline_w_reps_relu}}\label{sec:appendix_a}
\begin{proof}
    Let $\psi(x)$ denote the second order B-spline wavelet. We can express this as a linear combination of seven ReLU functions,
\begin{align}
\psi(x) = \frac{1}{6}\sigma(x) - \frac{8}{6}\sigma\left(x-\frac{1}{2}\right) + \frac{23}{6} \sigma(x-1) - \frac{16}{3} \sigma\left(x- \frac{3}{2}\right) + \frac{23}{6} \sigma(x-2) - \frac{8}{6}\sigma\left(x- \frac{5}{2}\right) + \frac{1}{6} \sigma(x-3).
\end{align}
With this representation of the B-spline wavelet, it is clear that one ReLU on the interval $[-1,1]$ can be represented in terms of a B-spline wavelet as
\begin{align}
    \sigma(x) \1_{[-1,1]} = 24 \psi\left(\frac{1}{4} x\right) \1_{[-1,1]},
\end{align}
where $\1_{[-1,1]}$ denotes an indicator function on $[-1,1]$. Therefore if we define $g: [-1,1] \to \R$ as
\begin{align}
    g(x) =  c + \sum_{k=1}^{K} 24 v_k \psi\left(\frac{1}{4} (w_k x - b_k)\right)
\end{align}
then clearly
\begin{align}
    f(x)= g(x) \quad \forall x \in [-1,1].
\end{align}
\end{proof}
\section{Proof of Theorem \ref{thm:bspline_w_cond}}\label{sec:appendix_b}
Our proof relies on Gershgorin's circle theorem which we recall here.
\begin{theorem}[Gershgorin circle theorem \cite{horn2012matrix}]
    Let $A \in \R^{n \times n}$ with entries $a_{ij}$. For $i = 1, \cdots, n$ let $R_i$ be the sum of the absolute values of off diagonals in each row of $A$
    \begin{align*}
        R_i(A) = \sum_{i \neq j} |a_{ij}|.
    \end{align*}
    Consider the $n$ Gershgorin discs defined as
    \begin{align*}
        \{z \in \mathbb{C}: |z-a_{ii}| \leq R_i(A)\}.
    \end{align*}
    Then every eigenvalue of $A$ lies within at least one of the Gershgorin discs.
\end{theorem}

\noindent We now proceed to the proof of \Cref{thm:bspline_w_cond}.
\begin{proof}[Proof of \Cref{thm:bspline_w_cond}]
    Our proof follows by bounding the sum of the absolute values of the off-diagonals of each row of $\mathbf{G}_{\psi}$ and then employing Gershgorin's circle theorem. Recall that,
    for each scale $j = 0,\dots, J-1$ we have $k=0,1,\dots, 2^{j}-1$ shifted versions of the neurons at this scale.
    For ease of notation we define the neurons as
    \begin{align}
        \psi_{j,k}(x) \coloneqq 2^{j/2} \cdot \psi(2^{j}(3/2)(x+1) - k).
    \end{align}
    Note that the scaling factor $2^{j/2}$ ensures that the $L_2$ norm of all the neurons are equal. The elements in $\mathbf{G}_{\psi} \in \R^{K \times K}$ consist of inner products between every pair of neurons.
    
    The diagonal entries of $\mathbf{G}_{\psi}$ are simply the squared $L_{2}$ norm of each neuron,
    \begin{align}
        \|\psi_{j,k}\|^2_2 &= \int_{D} (2^{j/2}\psi(2^j(3/2) (x+1)-k))^2 dx\\
        &= \int^{1}_{-1} (2^{j/2}\psi(2^j(3/2) (x+1)-k))^2 dx.
    \end{align}
    Let $t = \frac{3}{2} (x +1)$, then by a change of variables we have
    \begin{align}
        \|\psi_{j,k}\|^2_2 &=\frac{2}{3}\int^{3}_{0} (2^{j/2}\psi(2^{j}t-k))^2 dt
        = \frac{2}{3} \left(\frac{1}{4}\right) = \frac{1}{6}.
    \end{align}
    The $\frac{1}{4}$ follows from the fact that $\|\psi\|^2_2 = \frac{1}{4}$ and our normalization ensures that translations and dilation of the wavelet preserve its norm.  
    
    For the first row of $\mathbf{G}_{\psi}$ we have,
    \begin{align}
        R_1(\mathbf{G}_{\psi})= 0. 
    \end{align}
    This follows from the fact that each column in the first row consists of an inner product between $\psi_{0,0}$ and $\psi_{j,k}$ for all resolution of $j > 0$ and all possible shifts at each resolution. Since the B-spline wavelets are semiorthogonal \citep{chui1992compactly} we must have that whenever $i \neq j$ and for any $k,\ell \in \mathbb{Z}$,
    \begin{align}
        \langle \psi_{j,k} , \psi_{i, \ell} \rangle_{L_2(D)} = 0.
    \end{align}
 Where $\langle \cdot, \cdot \rangle_{L_2}$ denotes the $L_2$ inner product of two functions. From this we can conclude that $(\mathbf{G}_{\psi})_{0,\ell} = 0$ for all $\ell=0,\cdots, K-1$.

Now for the rest of the rows each row is identified with a neuron $\psi_{j,k}$ at a certain scale $2^j$ and shift $k$. The columns in this row are inner products of $\psi_{j,k}$ with all the other neurons in the network. These are the off-diagonals that we will bound.
For each neuron $\psi_{j,k}$ its inner product with all other neurons can be broken up into two cases. 
\paragraph{Case 1: The other neurons are at a different scale.}
The first case consists of inner products between other neurons which are at a different scale than $\psi_{j,k}$ i.e. $\langle \psi_{j,k}, \psi_{m,p} \rangle_{L_2}$ for $j \neq m$ but any $k,p$. Again, thanks to the semiorthogonality of the B-spline wavelets \citep{chui1992compactly} we have
\begin{align}
    \langle \psi_{j,k}, \psi_{m,p} \rangle_{L_2(D)} = 0.
\end{align}

\paragraph{Case 2: The other neurons are at the same scale.}
The second type of inner products are of the form,
\begin{align}
    \langle \psi_{j,k}, \psi_{j,p} \rangle_{L_2(D)},
\end{align}
where $k,p = 0,\dots, 2^{j}-1$ and $k \neq p$.
By the compactness of the B-spline wavelets, if $|k-p| \geq 3$ then the inner product is zero. Therefore it remains to bound 
\begin{align}
   \langle \psi_{j,k}, \psi_{j,p} \rangle_{L_2(D)}
\end{align}
for the case when $|k - p| = 1$ or $|k - p| = 2$. A direct computation, again applying a change of variables, reveals that for the case of $|k - p| = 1$ we have
    \begin{align}
        \frac{2}{3}\int^{3}_{0} 2^{j/2} \psi(2^j t) \cdot 2^{j/2}\psi(2^j t - 1) dt = 0.030864 = C_1.
    \end{align}
    Furthermore, when $|k - \ell| = 2$ another computation reveals that
    \begin{align}
       \frac{2}{3} \int^{3}_{0} 2^{j/2} \psi(2^j t) \cdot 2^{j/2} \psi(2^j t -2) dt=-0.0030864 = C_2.
    \end{align}

    At each scale $j$ we could have at most two neuron functions which are at the same scale $j$ but shifted by $1$ and at most two neuron functions which are at the same scale but shifted by $2$. 
    Therefore, by Gershgorin's circle theorem we have that for all the eigenvalues of $\mathbf{G}_{\psi}$ they must satisfy,
    \begin{align}
        \lambda_i \in \left\{z \in \R : \left|z - \frac{1}{6}\right| \leq 2\left(|C_1| + |C_2|\right)\right\}.
    \end{align}
    where $2(|C_1| + |C_2|) < \frac{1}{6}$.
    Therefore,
    \begin{align}
        \kappa(\mathbf{G}_{\psi}) = \mathcal{O}(1).
    \end{align}
\end{proof}
\section{Further Experiments on the Variation Norm}
\begin{figure*}[!th]
     \centering
    \begin{subfigure}[b]{0.24\columnwidth}
         \centering
         \includegraphics[width=\columnwidth]{figures_superres/butterfly_hr.pdf}
         \caption*{Original \\Full Resolution Image}
     \end{subfigure}
     \begin{subfigure}[b]{0.24\columnwidth}
         \centering
         \caption*{$\sum_{\ell}\|g^{(\ell)}_{\bm{\theta}}\|_{\mathcal{V}} = 2304$}
         \includegraphics[width=\columnwidth]{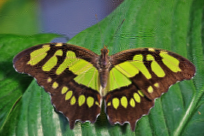}
         \caption{PSNR: 26.25 dB\\$c=2$}
     \end{subfigure}
     \hfill
     \begin{subfigure}[b]{0.24\columnwidth}
         \centering
         \caption*{$\sum_{\ell}\|g^{(\ell)}_{\bm{\theta}}\|_{\mathcal{V}} = 1358$}
         \includegraphics[width=\columnwidth]{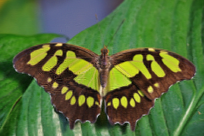}
         \caption{PSNR: \textbf{27.01 dB}\\$c=3$}
     \end{subfigure}
     \hfill
      \begin{subfigure}[b]{0.24\columnwidth}
         \centering
         \caption*{$\sum_{\ell} \|g^{(\ell)}_{\bm{\theta}}\|_{\mathcal{V}} = 1468$}
         \includegraphics[width=\columnwidth]{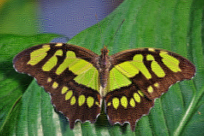}
         \caption{PSNR: 25.29 dB\\$c=5$
         }
     \end{subfigure}
    \caption{Three BW-ReLU DNNs trained on the superresolution task with different values of $c$. The $c$ which produces the highest PSNR corresponds to the one with the lowest variation norm across all layers.}
    \label{fig:superres_path_norm}
\end{figure*}

\begin{figure*}[!h]
     \centering
      \begin{subfigure}[b]{0.24\columnwidth}
         \centering
         \includegraphics[width=\columnwidth]{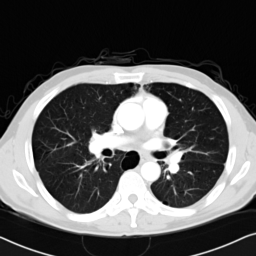}
         \caption*{Original\\ CT
         }
     \end{subfigure}
     \hfill
    \begin{subfigure}[b]{0.24\columnwidth}
         \centering
         \caption*{$\sum_{\ell}\|g^{(\ell)}_{\bm{\theta}}\|_{\mathcal{V}} = 1147$}
         \includegraphics[width=\columnwidth]{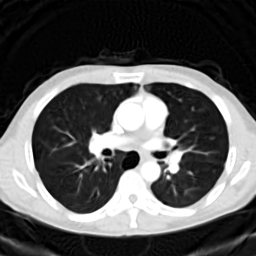}
         \caption{PSNR: \textbf{33.9 dB} \\$c=2$}
     \end{subfigure}
     \hfill
     \begin{subfigure}[b]{0.24\columnwidth}
         \centering
         \caption*{$\sum_{\ell}\|g^{(\ell)}_{\bm{\theta}}\|_{\mathcal{V}} = 1191$}
         \includegraphics[width=\columnwidth]{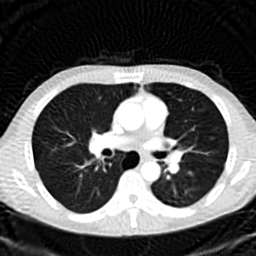}
         \caption{PSNR: 32.5 dB\\$c=3$}
     \end{subfigure}
     \hfill
     \begin{subfigure}[b]{0.24\columnwidth}
         \centering
         \caption*{$\sum_{\ell}\|g^{(\ell)}_{\bm{\theta}}\|_{\mathcal{V}} = 1601$}
         \includegraphics[width=\columnwidth]{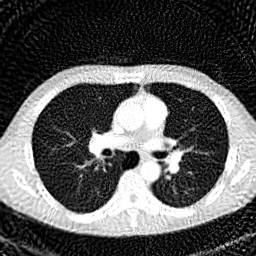}
         \caption{PSNR: 26.1 dB\\$c=5$}
     \end{subfigure}
    \caption{Three BW-ReLU DNNs trained on a different CT reconstruction task with different values of $c$. All networks are trained to the same training loss. Again the $c$ which produces the highest PSNR corresponds to the one with the lowest variation norm across all layers.}
    \label{fig:ct2_path_norm}
\end{figure*}

In this section we provide two more experiments demonstrating the correspondence between the optimal $c$ parameter and the network with the smallest variation norm across all layers. Our first experiment demonstrates that the same observation holds for the superresolution experiment in \cref{sec:experiments}. The results are shown in \Cref{fig:superres_path_norm}. We also repeated the experiment in \Cref{fig:ct_path_norm} on an alternative CT image. The results are shown in \Cref{fig:ct2_path_norm}.

\section{Training Details for Experiments}\label{sec:appendix-c}
For all experiments we applied an exponential learning rate scheduler of the form
\begin{align*}
    \eta(t) = \eta_0 (r)^{t/T},
\end{align*}
where $\eta_0$ is the initial learning rate, $T$ is the total epochs we trained for and $r$ is the decay rate.

\subsection{CT Reconstruction}\label{app:ct_dets}
We computed 100 equally spaced CT measurements of the X-ray image in the Radon domain. The DNN was trained by computing the mean squared error between the Radon transform of the image generated by the DNN and the ground truth CT measurements.
For the BW-ReLU neural network we used a learning rate of $\eta_0=2e-3$ with $c = 3$. For SIREN we trained with a learning rate of $\eta_0=1e-3$ and set $\omega_0 =25$. For WIRE a learning rate of $\eta_0 = 5e-3$ was used setting $\omega_0=7$ and $\sigma_0=10$. Finally for the ReLU + positional encoding we trained with a learning rate of $\eta_0 = 3e-3$. All methods were trained for 10000 epochs and the decay rate for the learning rate scheduler was $r=0.1$. In all cases the architecture consisted of 3 hidden layers and 300 neurons per layer.

\subsection{Signal Representation}\label{app:sigrep_dets}
For the BW-ReLU DNN we used an initial learning rate of $\eta_0 = 4e-3$ a scaling parameter applied to each neuron of $c=9$. For SIREN we trained with an initial learning rate of $\eta_0 = 2e-3$ and set the scale $\omega_0=50$. For WIRE we trained with a learning rate of $\eta_0 = 1e-3$ setting $\sigma_0=10$ and $\omega_0=20$. Finally for ReLU + positional encoding we used an initial learning rate of $\eta_0 = 4e-3$. All models were trained for 1000 epochs and the decay rate for the learning rate was $r=0.1$. The architecture consisted of 3 hidden layers and 300 neurons per layer.

\subsection{Super Resolution}\label{app:supres_dets}
We performed super resolution by training the DNN to minimize the mean squared error between a low resolution image of the Butterfly and a downsampled version of the full sample image produced by the DNN.
For the BW-ReLU DNN we trained with an initial learning rate of $\eta_0 = 3e-3$ and scaling parameter $c = 3$. For SIREN we used an initial learning rate of $\eta_0 = 2e-3$ and set $\omega_0=12$. For WIRE we used an initial learning rate of $\eta_0 = 3e-3$ with $\omega_0=8$ and  $\sigma_0=6$.  The ReLU + positional encoding architecture used an initial learning rate of $\eta_0 = 4e-3$. All models were trained for 2000 epochs and the decay rate for the learning rate scheduler was $r=0.2$. In all cases the architecture consisted of 3 hidden layers and 256 neurons per layer.

\subsection{Low Norm Solutions and Inverse Problems}\label{app:low_norm_dets}
The learning rate we used in this experiment was varied for each value of $c$ to ensure that all the networks achieved equal training loss. We considered the DNN as a composition of three shallow BW-ReLU neural networks and summed up the variation norm of each layer according to \eqref{eq:variation_norm_bspline_c}. In particular for the architecture defined as,
\begin{align}
    g(\vx) = \mathbf{W}_3 \psi\left(c \cdot \mathbf{W}_2 \psi \left(c \cdot \mathbf{W}_1 \psi \left(c \cdot \mathbf{W}_0 \vx\right)\right)\right).
\end{align}
We can treat it as a composition of 3 shallow nets of the form

\begin{align}
    g^{(1)}(\vx) &= \mathbf{W}_{1} \psi\left(c \cdot \mathbf{W}_0 \vx \right) \quad \vx \in \R^2 \\
    g^{(2)}(\vx) &= \mathbf{W}_2 \psi\left( c \cdot \mathbf{I}\vx\right) \quad \vx \in \R^{K}\\
    g^{(3)}(\vx) &= \mathbf{W}_3 \psi(c \cdot \mathbf{I}\vx) \quad \vx \in \R^{K}
\end{align}
and then the whole DNN is
\begin{align}
    g(\vx) = g_3 \circ g_2 \circ g_1.
\end{align}
The sum of the variation norm across all layers would be 
\begin{align}
    \sum_{\ell} \|g^{(\ell)}\|_{\mathcal{V}} = &16c\left(\sum_{k}^{K} \|\vw^{(1)}_k\|_2 \|\vw^{(0)}_k\|_2 +\sum_{k=1}^{K} \|\vw_{k}^{(2)}\|_2+\sum^{K}_{k=1} \|\vw_{k}^{(3)}\|_2\right).
\end{align}

\end{document}

%% file: math_commands.tex
\usepackage{amsmath,amsfonts,bm}
\usepackage{bbm}
\usepackage[mathscr]{eucal}

\def\1{\bm{1}}

\def\<{\langle}
\def\>{\rangle}

\def\vr{{\bm{r}}}

\def\vv{{\bm{v}}}
\def\vw{{\bm{w}}}
\def\vx{{\bm{x}}}
\def\vy{{\bm{y}}}

\DeclareMathAlphabet{\mathsfit}{\encodingdefault}{\sfdefault}{m}{sl}
\SetMathAlphabet{\mathsfit}{bold}{\encodingdefault}{\sfdefault}{bx}{n}

\newcommand{\R}{\mathbb{R}}

